\documentclass{article}
\usepackage[utf8]{inputenc}
\usepackage{amsmath}
\usepackage{graphicx}
\usepackage[square,numbers]{natbib}
\usepackage{color}
\usepackage[a4paper, total={6in, 8in}]{geometry}
\usepackage{authblk}
\usepackage{verbatim}
\usepackage{fancyhdr}
\title{Graphene-Gold THz Metasurfaces with Tailored Resonant Structure for Enhanced Nonlinear Response}

\author[1,2]{A.~Theodosi}
\author[3]{I.~A.~Otoo}
\author[1,4]{A.~D.~Koulouklidis}
\author[5]{N.~Matthaiakakis}
\author[6]{G.~Kakarantzas}
\author[7]{P.~Mustonen} % Petri
\author[7]{H.~Lipsanen} % Harri
\author[3]{G.~Fedorov} %Georgy
\author[1]{I.~Liontos}
\author[1,2]{S.~Tzortzakis}
\author[3]{P.~Kuzhir}
\author[1,2]{M.~Kafesaki}
\author[6,1]{O.~Tsilipakos\thanks{otsilipakos@eie.gr}}

\affil[1]{Institute of Electronic Structure and Laser, Foundation for Research and Technology Hellas, 70013 Heraklion, Greece}
\affil[2]{Department of Materials Science and Engineering, University of Crete, 70013 Heraklion, Greece}
\affil[3]{Department of Physics and Mathematics, University of Eastern Finland, 80101 Joensuu, Finland}
\affil[4]{Regensburg Center for Ultrafast Nanoscopy (RUN) and Department of Physics, University of Regensburg, 93040 Regensburg, Germany}
\affil[5]{Institute of Nanoscience and Nanotechnology, National Center for Scientific Research “Demokritos”,
15341 Athens, Greece}
\affil[6]{Theoretical and Physical Chemistry Institute, National Hellenic Research Foundation, 11635 Athens, Greece}
\affil[7]{Dept. Electronics and Nanoeng. Aalto University, Maarintie 8, 02150 Espoo, Finland}

\begin{document}

\maketitle

\begin{abstract}
Graphene’s exceptional nonlinear optical properties combined with resonant photonic structures offer a promising pathway for efficient nonlinear applications at terahertz (THz) frequencies. In this work, we propose and demonstrate a fabrication-friendly hybrid nonlinear metasurface composed of gold patches integrated with uniform graphene, circumventing the need for complex graphene patterning. The structure supports strong localized resonances that enhance nonlinear interactions. By exploiting resonant enhancement at both the fundamental and third harmonic frequencies, we  predict via simulations third-harmonic generation efficiencies as high as -15 dB (3.2\%) under continuous-wave excitation at modest intensities (0.1 MW/cm$^2$). The metasurface is fabricated via electron-beam lithography and experimentally characterized using THz time-domain spectroscopy. Under pulsed excitation, we experimentally observe pronounced nonlinear frequency shifts up to 0.5~THz (12.5\% fractional change), driven by self-phase modulation, consistent with simulation results. These findings highlight the potential of tailored graphene-based metasurfaces for efficient nonlinear THz photonic devices.
%Graphene is a unique two dimensional material for nonlinear applications in the THz regime, due to its high third order nonlinearity and the ability to support tightly confined surface plasmons. In this work, we propose a hybrid gold-patch/uniform-graphene metasurface implementation that is friendlier for fabrication compared to patterned graphene approaches and, at the same time, aims to further increase the efficiency of the third harmonic generation process. The efficiency of the nonlinear process is enhanced by spectrally aligning the fundamental and third harmonic frequencies with resonances of the metasurface, leading to spatiotemporal energy confinement in both steps, of excitation at $\omega$ and radiation at $3\omega$. Conversion efficiencies as high as -15~dB (3.2\%) for an input intensity of 0.1~MW/cm$^2$ are theoretically predicted for continuous-wave operation. The nonlinear metasurface is fabricated and measured by a THz time-domain spectroscopy setup under pulsed excitation. Strong nonlinear frequency shifts up to 0.5~THz due to self-phase modulation are experimentally observed and verified by corresponding simulations. Our results highlight the potential of graphene-based metasurfaces with a tailored resonance structure for nonlinear applications.
\end{abstract}

\section{Introduction}
Metasurfaces (MSs), ultrathin structures composed of a periodic arrangement of resonant meta-atoms on a plane, have attracted considerable research interest \cite{QuevedoTeruel2019}. They have been shown to successfully control the amplitude \cite{Liu:2017}, phase \cite{Tsilipakos2023},  wavefront \cite{Tsilipakos:2020aom}, and polarization \cite{Orazbay2024} of incident radiation. In order to enable nonlinear and reconfigurable MSs, materials with intensity-dependent and tunable properties can be exploited in the meta-atom geometry \cite{Savarese2025,DeglInnocenti2025,Gu2025}. Nonlinearity, in particular, can significantly broaden the scope of achievable functionalities, enabling all-optical control and the generation of new frequencies. In  this context, a unique nonlinear material for nonlinear MSs at THz frequencies is graphene \cite{Vermeulen2022}, which combines strong third order nonlinearity \cite{Mikhailov2019,Dremetsika2016}, the ability to dynamically tune its linear and nonlinear properties via gating \cite{Alexander2018,Amanatiadis}, and the capability to support tightly-confined surface plasmons. 
%Importantly, the short wavelength of propagating graphene plasmons enables the design of metasurfaces (MSs) supporting higher-order resonances, while the lattice constant remains subwavelength even for higher harmonic frequencies, avoiding diffraction effects.

%In addition, nonlinear metasurfaces, made of nonlinear materials, enable extra degrees of freedom and functionalities such as generating new frequencies \cite{Jin2017, Christopoulos2018, Theodosi2021}. 

The potential of graphene as a nonlinear material has been lucidly demonstrated in \cite{Hafez2018}, for instance, where high harmonic generation (up to seventh order) was obtained by a uniform graphene layer on a glass substrate. Patterning graphene into a metasurface can allow for fostering resonances, which enhance its interaction with light. By exploiting the corresponding energy confinement in space (small effective mode volume) and time (high quality factor), the efficiency of the nonlinear process can be further enhanced. This strategy has been leveraged for proposing efficient third harmonic generation \cite{Jin2017,You2017,Theodosi2021} and all-optical polarization control \cite{Matthaiakakis2024}, among others. However, the patterning of graphene is an experimentally demanding procedure. To circumvent this difficulty, we propose here an experimentally friendlier (but still equally efficient) implementation based on a uniform graphene layer, where the patterning is ``transferred'' to an adjacent gold layer. This approach still allows us to achieve strong resonances and tailor the resonant structure but without patterning graphene. 

Specifically, here we propose a hybrid gold-patch/uniform-graphene metasurface for enhancing nonlinear phenomena at THz frequencies. By varying the dimensions of the gold patches we can finely tune the resonant frequencies. It, thus, becomes possible to position two well-defined resonances at frequencies with ratio 1:3. Aligning resonances of the metasurface with the fundamental (FF) and third-harmonic (TH) frequencies is a well-known strategy for enhancing the conversion efficiency (CE) of the third harmonic generation (THG) process and is frequently termed doubly-resonant enhancement \cite{You2017}. Conversion efficiencies as high as -15~dB (3.2\%) for an input intensity of 0.1~MW/cm$^2$ are theoretically predicted for continuous-wave operation, which are superior to previous approaches based on patterned graphene metasurfaces \cite{Jin2017, Theodosi2021}. Subsequently, the proposed metasurface is fabricated and measured with a THz time-domain spectroscopy setup under pulsed excitation. Strong nonlinear frequency shifts up to 0.5~THz due to self-phase modulation are experimentally observed and verified by corresponding simulations. This pronounced nonlinear shift corresponds to a fractional decrease of the resonant frequency by 12.5\% and can be used for nonlinearly switching between values of low and high reflection by tuning the input intensity.

\section{Results\label{sec:results}}

\subsection{Theoretical Analysis and Design\label{sec:theory}}

%To illustrate the evolution of the design process, we start form a simple metal-backed substrate, which supports Fabry-Perot resonances. The addition of graphene allows for increasing the absorption and forming well-defined spectral resonant features. We design for positioning two resonances which are compatible with the source. A simple structure of uniform graphene on a metal backed substrate is not adequate. 

%\red{We want to tailor the resonant structure into achieving strong resonances. We target two resonances in order to be able to resonantly enhance two frequencies, as for example with the THG process.}

%This section proposes a tunable multi-resonant hybrid gold-graphene metasurface design for efficient nonlinear response. We describe the process of the design procedure and we present the parametric study based on the optimization of the THG conversion efficiency, resulting in the optimum geometrical parameter set offering the maximum efficiency due to the alignment of two resonances at $\omega$ - 3$\omega$ configuration. We indentify beyond doubt that these are the two resonances that mediate the nonlinear conversion process, by comparing their eigenvectors with the field distribution that is observed in the processes of illumination at $\omega$ and radiation at $3\omega$.

%\subsection{Design evolution}

\begin{figure}[]
\centering
\includegraphics[width=15.2cm]{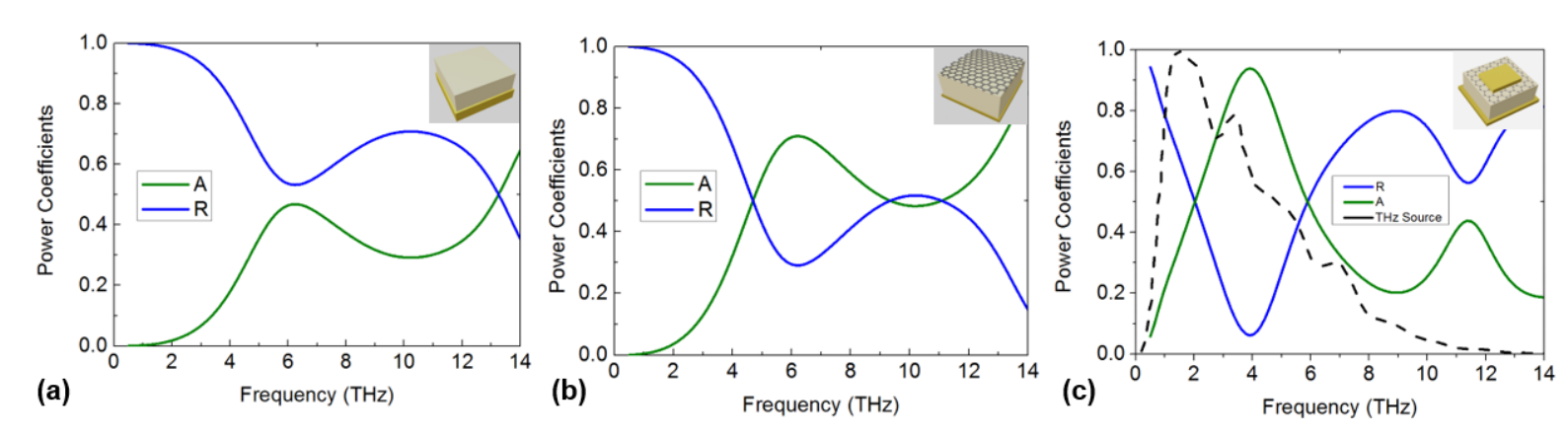}
\caption{Design procedure of hybrid gold-graphene metasurface. (a)~By placing a gold back-reflector in an 8.3-$\mu$m-thick substrate, broad Fabry-Perot resonances are observed in the reflection ($R$) and absorption ($A$) power coefficients. (b)~A uniform graphene layer on top of the gold-backed substrate leads to enhanced absorption. (c)~Incorporating gold meta-atoms allows to excite graphene surface plasmons, and two well-defined resonances appear within the frequency range of interest. Proper selection of the gold patch dimensions allows to position the resonances at an $\omega$ - 3$\omega$ configuration, which is desirable for doubly-resonant enhancement of the third harmonic generation process. The dashed line in (c) shows the power spectrum of the THz source employed in the experimental characterization.
\label{fig:DesignEvo}}
\end{figure}

In Fig.~\ref{fig:DesignEvo}, the evolution of the design process for the hybrid gold-graphene metasurface is presented. Initially, by placing a gold back-reflector in an 8.3-$\mu$m-thick substrate, a broad Fabry-Perot resonance is observed [Fig.~\ref{fig:DesignEvo}(a)]. Incorporating a uniform graphene layer on top of the gold-backed substrate leads to enhanced absorption [Fig.~\ref{fig:DesignEvo}(b)]. However, in both cases a single resonance (absorption peak) appears in the frequency range of interest (0-14~THz). The integration of gold patches allows to excite surface plasmons on graphene (otherwise there is a momentum mismatch with free-space photons), leading to two well-defined resonances within the frequency window 0-14~THz [Fig.~\ref{fig:DesignEvo}(c)]. By varying the shape and size of the gold meta-atoms, we can tune the resonant frequencies.  For a specific combination of parameters, the two resonances are perfectly aligned at an $\omega$ - 3$\omega$ configuration, which is the required condition to resonantly enhance both processes, of excitation at $\omega$ and radiation at $3\omega$. The desired positioning of the resonant frequencies is also related with the frequency content of our experimentally available input pulse [see dashed line in panel Fig.~\ref{fig:DesignEvo}(c)]. More specifically, the goal is to place the first resonance close to the peak of the pulse spectrum so as to best exploit the available field strength and allow the nonlinear processes to manifest. At the same time, we want to avoid exciting the second resonance with the input pulse, so that any recorded radiation at that frequency is solely attributed to the nonlinear harmonic generation process. We try to satisfy both constraints by placing the two frequencies at $\sim4$ and $\sim12$~THz, respectively.

%\subsection{Analysis and design for efficient third harmonic conversion}

\begin{figure}[]
\centering
\includegraphics[width=15cm]{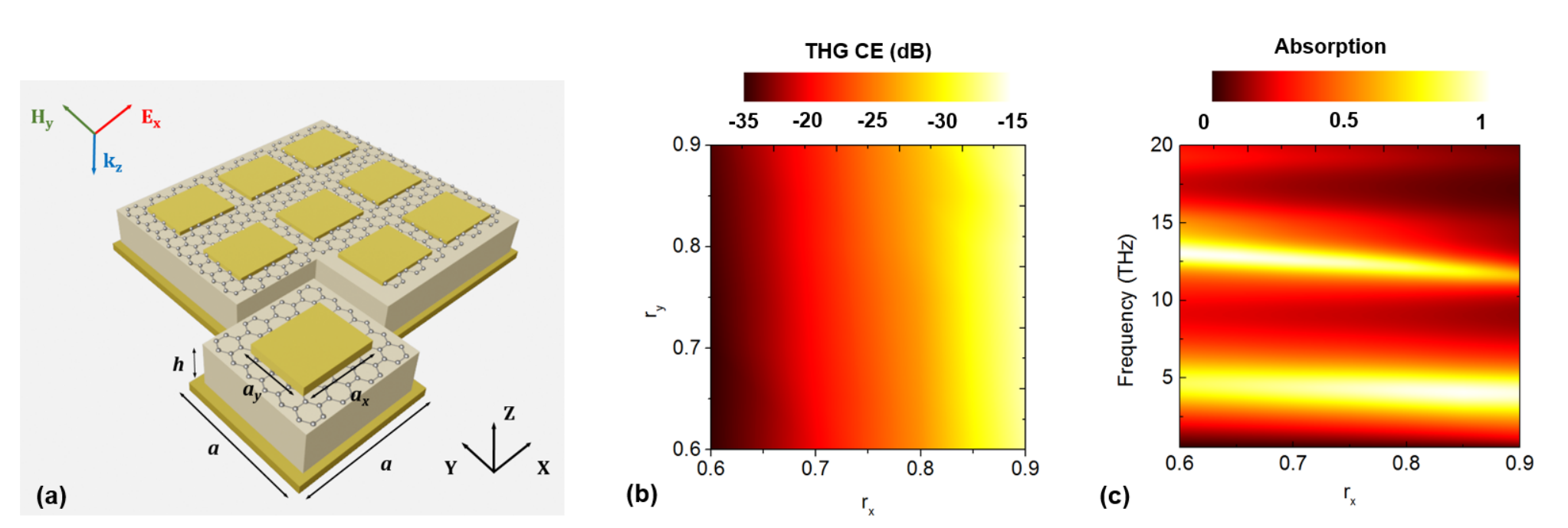}
\caption{(a) Schematic of the proposed hybrid gold-graphene metasurface. The metasurface is composed of rectangular gold patches $(a_x \times a_y)$ with thickness $t = 500$~nm and electric conductivity $\sigma = 4 \times 10^7$~S/m, placed in a square periodicity with period $a = 5.6~\mu$m. The patches reside  on a graphene-covered polymer substrate (relative electric permittivity $\epsilon_r = 3.2 \times(1-0.1j)$ and thickness $h = 8.3~\mu$m), backed by a gold back reflector. The graphene layer is uniform, which helps with fabrication simplicity. The electric field of the incident wave is linearly polarized along the $x$-axis. (b)~Calculated THG conversion efficiency (CE) under continuous plane wave illumination at normal incidence for an input intensity of $I = 0.1$~MW/cm$^2$, as a function of the gold patch size: $r_x = a_x/a$ and $r_y = a_y/a$. (c) Calculated linear absorption of the hybrid graphene metasurface, as a function of the frequency and patch size: $a_x = r_x \times a$ ($r_y$ is constant at $0.65$).  The absorption peaks of the two resonances are placed at $\sim4$ and $\sim12$~THz for $r_x = 0.9$. \label{fig:Parametric}}
\end{figure}

Through this design evolution process we reach the metasurface geometry depicted in Fig.~\ref{fig:Parametric}(a), where the main dimensions are also annotated. To arrive to this geometry we first make rational choices for the main unit cell dimensions: (i)~the lattice constant is selected such that the array remains subwavelength even at the third harmonic frequency in order to avoid diffraction effects \cite{Christopoulos:2024prb}, and (ii)~the substrate thickness is tuned so that the fundamental resonance is placed near 4~THz, as discussed in the previous paragraph. The next step is to conduct a parametric study regarding the dimensions of the gold-meta-atoms to tune the alignment of the resonances at the $\omega$ - 3$\omega$ configuration and to ensure the optimum performance. 
To enable extra degrees of freedom, we allow for a rectangular patch geometry, meaning that the metasurface response will differ for the two orthogonal linear polarizations. Unless otherwise noted, the incident wave is a normally-incident linearly polarized ($E_x$) plane wave, as depicted in the inset of Fig.~\ref{fig:Parametric}(a).
This optimization process is based on the maximization of the calculated third harmonic generation. The simulations are conducted using the commercial software COMSOL Multiphysics under continuous wave conditions by utilizing two independent frequency-domain linear simulations. Details can be found in Appendix~A. The gold meta-atoms with size $a_x \times a_y$, thickness $t = 500$~nm and electric conductivity $\sigma = 4 \times 10^7$~S/m, are placed in a square periodicity with period $a = 5.6~\mu$m, on top of a polymeric substrate with relative electric permittivity $\epsilon_r = $3.2 $\times$(1 – 0.1j) and thickness $h = 8.3~\mu$m. A unit cell of the metasurface (periodic boundary conditions at the $xz$ and $yz$ planes) is used in the simulations. Graphene is modeled using appropriate dispersive linear and third-order nonlinear conductivities \cite{Cheng2014, Jin2017} via the surface current boundary condition. The gold backreflector is significantly thicker than the skin depth at low THz frequencies; thus, transmission is hindered and the metasurface operates in reflection.

For the parametric study, we vary $r_x$ and $r_y$ ($r_x = a_x/a$ and $r_y = a_y/a$) in the range $0.6-0.9$ and calculate the corresponding THG efficiency [Fig.~\ref{fig:Parametric}(b)]. The input intensity is set to $I=0.1$~MW/cm$^2$ and the THG conversion efficiency (CE) was defined as the ratio of power radiated at the third harmonic to the input power at the fundamental frequency (P$_\mathrm{TH}/$P$_\mathrm{FF}$). In each case, the third harmonic frequency was chosen to match the second (and narrower) absorption peak of the metasurface [see Fig. \ref{fig:DesignEvo}(c)], while the operating (or fundamental) frequency is one third of this value. The maximum CE was calculated to be -15~dB ($\sim3.2\%$) for the combination $r_x=0.9$ and $r_y=0.65$.  To verify that the physical origin of this high efficiency is the doubly-resonant enhancement, in Fig.~\ref{fig:Parametric}(c) we plot the linear absorption of the hybrid graphene metasurface as a function of the frequency and patch size: $a_x = r_x \times a$. The size of the patch along the $y$-axis is held constant at $r_y = 0.65$.  Results for other values of the parameter $r_y$ are presented in Appendix~B. As shown in Fig.~\ref{fig:Parametric}(c), the two resonances are placed at frequencies $\sim4$ and $\sim12$~THz that best satisfy the $\omega$ - 3$\omega$ relation for the value $r_x=0.9$.

%\subsection{Underlying resonant structure}

\begin{figure}[t!]
\centering
\includegraphics[width=15cm]{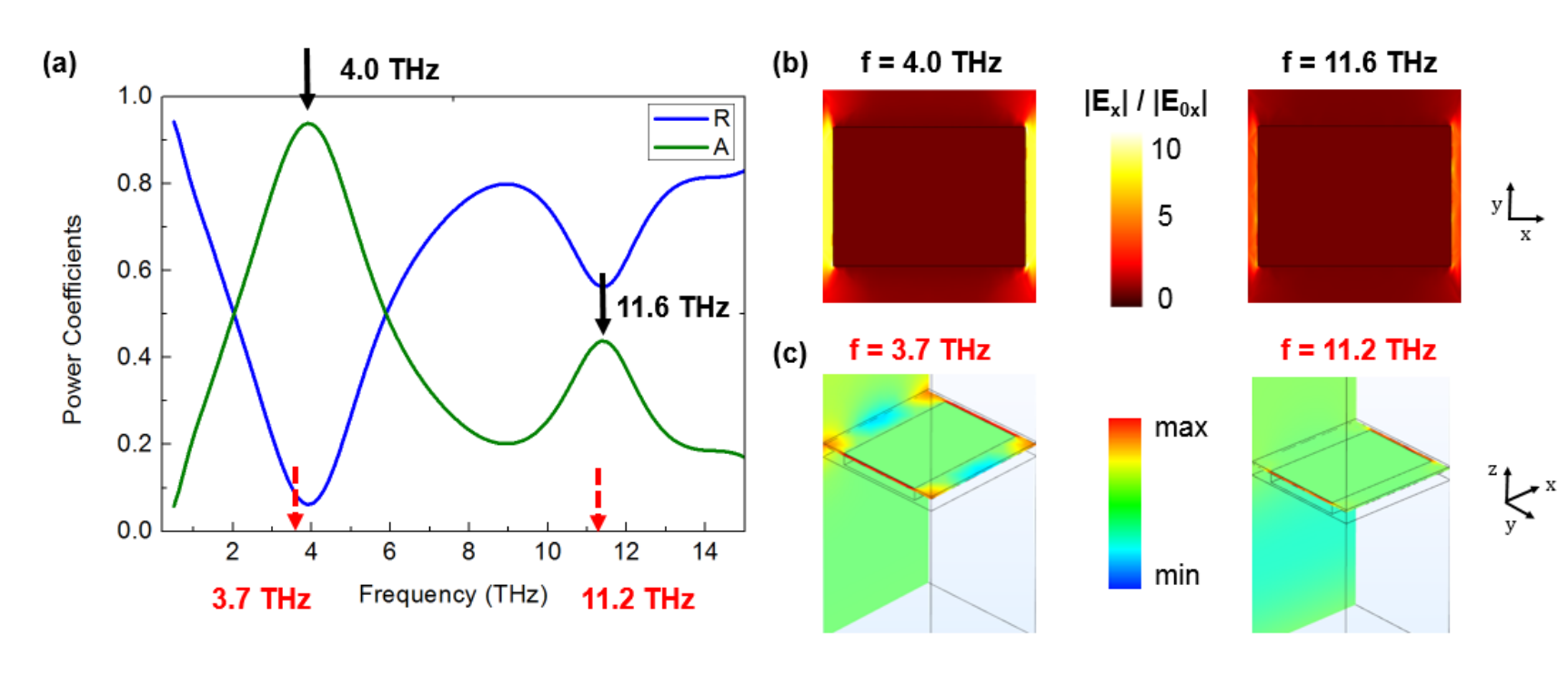}
\caption{Analysis of the optimum gold-graphene metasurface with $r_x = 0.90$ and $r_y = 0.65$. (a) Reflection ($R=|r|^2$) and Absorption ($A$) power coefficients of the metasurface obtained from linear plane-wave scattering simulations. The positions of the resonances are marked with black solid arrows and corresponding eigenfrequencies obtained from eigenmode analysis are marked with red dashed arrows. (b) Field enhancement factor $|E_x|/|E_{x0}|$, calculated at the plane of graphene. The maximum field enhancement factor is approximately 10 for the first resonance and 5 for the second. (c) Mode profile ($E_x$ component) of the two eigenmodes, shown in a perspective view of the metasurface (see axes). \label{fig:DoubleRes}}
\end{figure}

We can further study the electromagnetic response of the optimum metasurface ($r_x=0.9$ and $r_y=0.65$) by correlating eigenmode analysis and scattering simulations. In Fig.~\ref{fig:DoubleRes}(a), we plot the power coefficients for linear plane-wave scattering  (reflection: blue; absorption: green).  With black arrows we mark the positions of the two absorption peaks at 4~THz and 11.6~THz, respectively. At these frequencies, the field enhancement is maximized. By plotting the ratio $|E_x|/|E_{x0}|$ on the plane of graphene [Fig.~\ref{fig:DoubleRes}(b)], we find that the enhancement reaches a value of $\sim10$ for the first peak and $\sim5$ for the second one.  Finally, we use eigenmode analysis to calculate the eigenmodes of the metasurface. We verify that the two absorption peaks are directly associated with two eigenfrequencies of the metasurface at 3.7~THz and 11.2~THz, respectively. The eigenfrequencies are marked with red arrows in  Fig.~\ref{fig:DoubleRes}(a). The corresponding field distributions of the two eigenfrequencies are plotted in Fig.~\ref{fig:DoubleRes}(c).

\subsection{Experimental Demonstration\label{subsec:exp}}

In this Section, we proceed with the experimental demonstration.  We first describe step-by-step the fabrication process. Next, we present the material and structural characterization via Raman spectroscopy of the graphene monolayer and scanning electron microscope (SEM) images of the fabricated metasurface. Finally, we present the THz time-domain spectroscopy setup employed for the electromagnetic characterization and the corresponding measurements. 

\subsubsection{Fabrication and Inspection\label{subsec:Fabrication}}

\begin{figure}[]
\centering
\includegraphics[width=14cm]{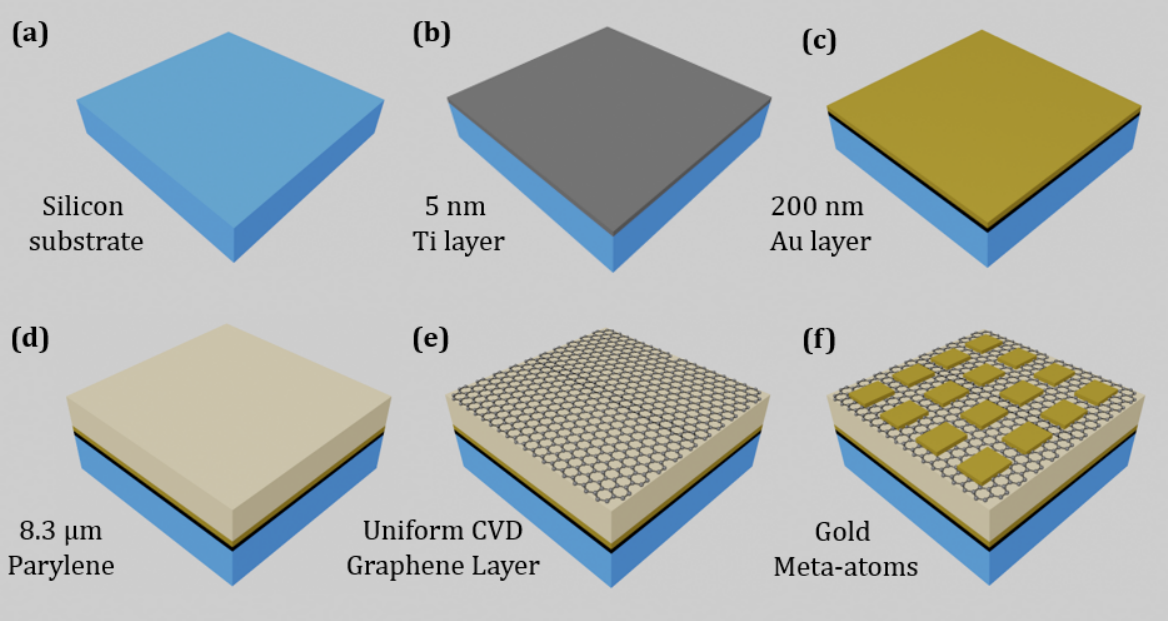}
\caption{Flow diagram of the fabrication process. (a)~A 2.5 cm $\times$ 2.5 cm piece of silicon was used as the substrate. (b)~A PVD evaporator evaporated a 5~nm thick titanium layer on top of silicon. (c)~Next, a 200~nm thick gold layer was evaporated on Ti and Si to form the backreflector. (d)~Addition of an 8.3-$\mu$m-thick parylene layer. (e)~A single CVD graphene layer was transferred on top of parylene. (f)~Addition of gold meta-atoms.
\label{fig:4}}
\end{figure}

The samples were fabricated using 525±25~$\mu$m thick single-sided polished 100 mm diameter, phosphorus and boron-doped silicon wafers with low resistivity 1-30 $\Omega$cm [Fig.~\ref{fig:4}(a)]. The Si wafers were first cleaved into 2.5 cm $\times$ 2.5 cm  (about 0.98 square inches) size and then  cleaned with acetone in an ultrasonic water bath for 5 minutes at 100\% power. Then, they were transferred into another beaker containing isopropanol and cleaned in an ultrasonic water bath for 5 minutes. The samples were air-dried by blowing nitrogen gas (N$_2$) on them. Next, the samples were  coated with 5 nm of titanium [Fig.~\ref{fig:4}(b)] and 200 nm of thermally evaporated gold on top of Ti [Fig.~\ref{fig:4}(c)]. The thermal evaporation was conducted using an Amod physical vapor deposition (PVD) device by Angstrom engineering, at Micronova lab, Aalto University.  On top of the evaporated Au, a layer of parylene was grown using a Specialty Labcoter 2 Parylene deposition system device (PDS 2010) operated at 650~$^\circ$C for evaporation and at room temperature for deposition. The final thickness of the parylene layer was determined by an ellipsometer and a profilometer and was found to be 8.3~$\mu$m. Subsequently, a single graphene layer was transferred on parylene as shown in [Fig.~\ref{fig:4}(e)]. The final step was to prepare and place the gold rectangular meta-atom array on the samples. Firstly, a positive electron beam resist (AR-P 671.11) was spin-coated on the substrate, followed by baking the substrate on a hot plate at 150~$^\circ$C for 5 minutes. The resist was exposed using the Raith EBPG 5000+ ESHR electron beam lithography system and then developed using a standard in-house developer. Next, the samples were coated with 5~nm of chromium (Cr) for adhesion, and 200 nm of Au was evaporated on the Cr using a MiniLab 026 evaporator device. After the thermal evaporation of gold, the sample was put in a beaker containing acetone for 12 hours to lift the excess gold around the meta-atoms and the substrate. Finally, the samples were transferred to a beaker containing isopropanol for 5 minutes to wash away all the acetone. The resulting samples of the hybrid graphene-gold metasurfaces [Fig.~\ref{fig:4}(f)] were gently dried by blowing dry nitrogen gas.

\begin{figure}[]
\centering
\includegraphics[width=15.2cm]{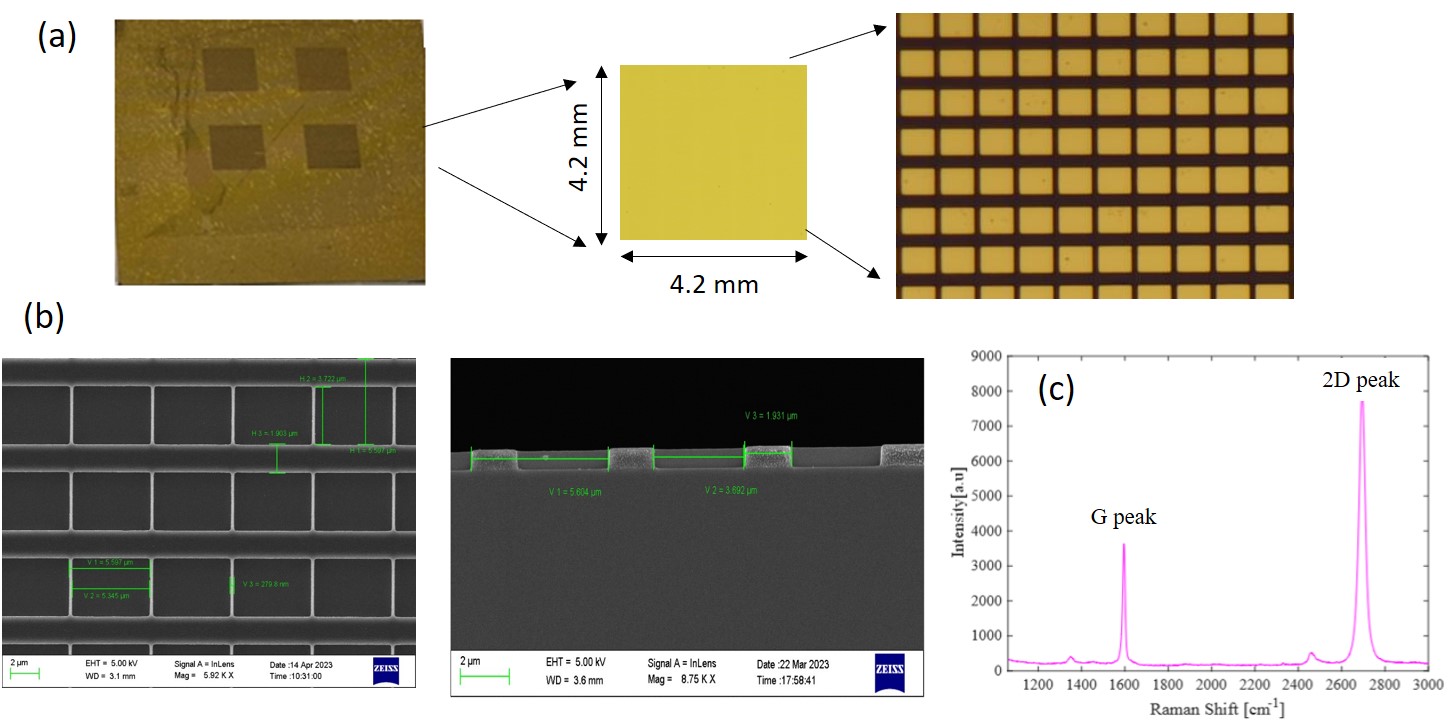}
\caption{(a) Optical image of the complete metasurface with lateral dimensions of 4.2 mm $\times$ 4.2 mm. A magnified view illustrates the detailed configuration of the individual unit cells within the metasurface array. (b) SEM images of the fabricated graphene-based metasurface after development confirm the structural dimensions achieved during the fabrication process. The SEM images also show both the top view and the lateral view of the fabricated sample. (c)~Raman spectroscopy data of a single-layered graphene transferred onto the metasurface. The two characteristic graphene G (1595.4 cm$^-1$) and 2D (2695.8 cm$^-1$) peaks are presented. 
%(a-b) Inspection of the fabricated metasurface. Each metasurface is $\sim 4~\mathrm{mm}\times 4~\mathrm{mm}$. It comprises roughly 700 $\times$ 700 unit cells and amounts to an extent of 160 free space wavelengths at the third harmonic frequency (12~THz).
%(c)~Raman spectrum of transferred CVD graphene. Notice the two characteristic graphene G (1595.4 cm$^{-1}$) and 2D (2695.8 cm$^{-1}$) peaks. (d)~SEM image with annotated dimensions of gold meta-atoms. The lattice constant is 5.591~$\mu$m in perfect agreement with the the initial design (5.6~$\mu$m) verifying the accuracy of the manufacturing process.
\label{fig:5}}
\end{figure}

Optical images of the fabricated hybrid gold-graphene metasurfaces are presented in Fig.~\ref{fig:5}(a). Each metasurface is of total area $\sim 4.2~\mathrm{mm}\times 4.2~\mathrm{mm}$. It comprises roughly 700 $\times$ 700 unit cells and amounts to an extent of 168 free space wavelengths at the third harmonic frequency (12~THz). Raman spectroscopy was used to detect the two characteristic graphene G (1595.4 cm$^{-1}$) and 2D (2695.8 cm$^{-1}$) peaks [Fig.~\ref{fig:5}(c)]. Furthermore, SEM images are used to identify the exact shape and  dimensions of the fabricated rectangular gold-meta-atoms  [Fig.~\ref{fig:5}(b)]. The lattice constant is 5.597~$\mu$m in perfect agreement with the the initial design (5.6~$\mu$m) verifying the accuracy of the manufacturing process. In  the metasurface of Fig.~\ref{fig:5}(b), we were able to go up to a value $r_x=0.95$  with a well defined gap (280~nm). 

\subsubsection{THz Characterization with Time Domain Spectroscopy\label{subsub:THzChar}}

\begin{figure}[]
\centering
\includegraphics[width=15cm]{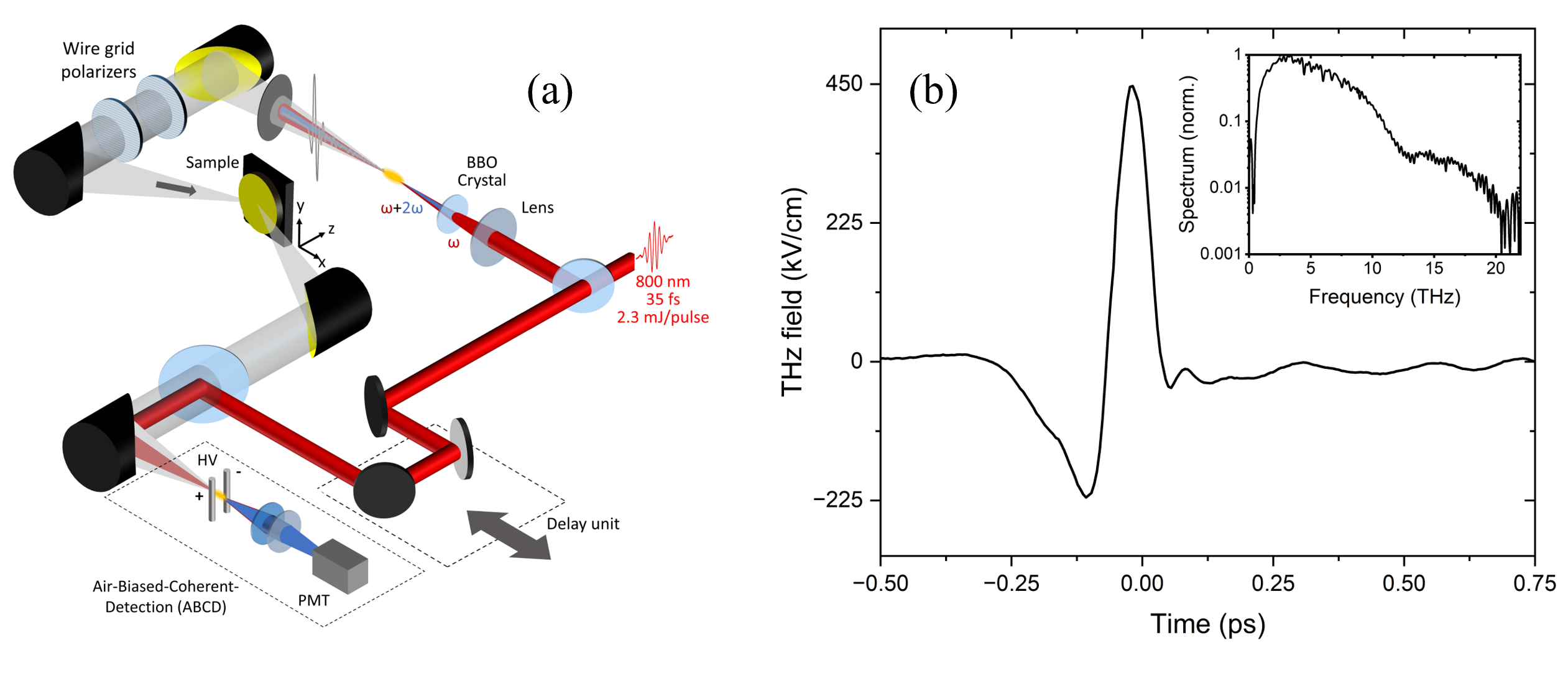}
\caption{(a)~Schematic of the experimental THz-TDS setup. (b)~Produced THz electric field indicating an ultra-short (roughly single-cycle) pulse. The corresponding spectrum is plotted in the inset. \label{fig:6}}
\end{figure}

The fabricated graphene-gold samples were electromagnetically characterized using a custom THz time-domain spectroscopy (THz-TDS) setup [Fig.~\ref{fig:6}(a)]. Broadband THz pulses with peak electric fields up to 450 kV/cm were generated via two-color filamentation of ultrashort laser pulses in air \cite{Koulouklidis2020-eh}. Figure~\ref{fig:6}(b)  displays the resulting THz electric field employed in the experiments, with the corresponding spectrum shown in the inset. The THz beam was subsequently directed through a pair of wire-grid polarizers, enabling intensity control, and then illuminated the sample under TM polarization at an incidence angle of 16$^\circ$ [Fig.~\ref{fig:6}(a)]. The reflected THz radiation was detected using air-biased coherent detection (ABCD), which provides access to the full spectral content of the THz pulse without interference from phonon resonances.

%The reflected THz waves are then driven to a 100-$\mu$m thick gallium phosphide (GaP) crystal for detection through electro-optic (EO) sampling. In this detection method, the THz beam co-propagates with a linearly polarized probe laser beam through GaP. The THz electric field changes the polarization ellipsoid of the refractive index in the EO crystal, inducing changes in the polarization of the probe pulse. These changes in the probe’s polarization are translated to intensity changes by an analyzer, e.g., by using a Wollaston prism. %\cite{Koulouklidis_2016}.
%A delay unit is used to control the delay between THz and probe beams, allowing for the transient detection of THz induced changes in the probe’s polarization, upon incidence in a pair of balanced photodiodes [Fig.~\ref{fig:6}(a)]. In the setup utilized for the experiments studied here, a 12 ps temporal window was achieved.

\begin{figure}[]
\centering
\includegraphics[width=15cm]{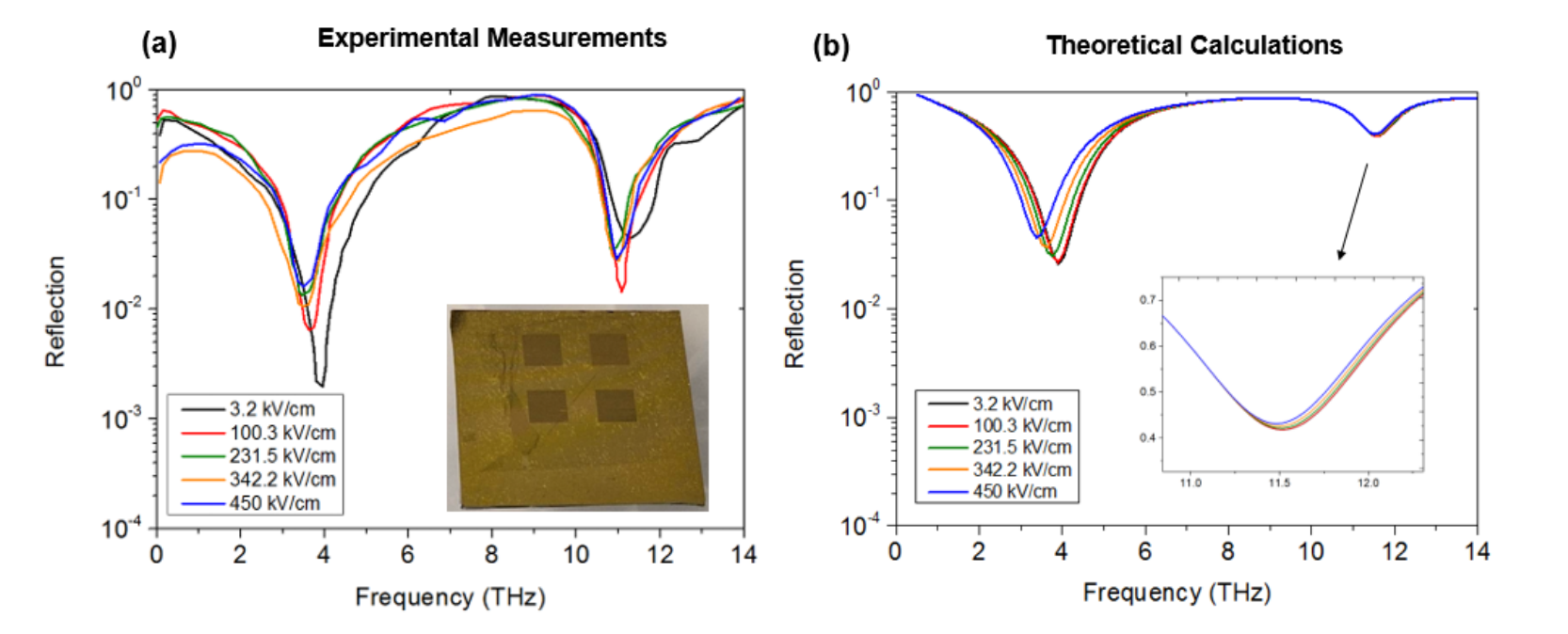}
\caption{(a) Experimental THz-TDS reflection spectra of the fabricated  hybrid gold-graphene metasurface (shown in the inset) for different values  of incident electric field strength. (b) Corresponding calculations (zoom in for the second resonance in the inset).\label{fig:7}}
\end{figure}

The experimentally-measured reflection spectra are presented in Fig.~\ref{fig:7}(a) as a function of the electric field amplitude $|E_0|$, varying from 3.2~kV/cm up to 450~kV/cm. The two reflection dips near 4 and 12~THz are clearly visible. Evidently, increasing the input intensity leads to a redshift of both resonances. This nonlinear self-shift is due to self-phase modulation; a $\Delta\omega<0$ shift is anticipated for self-focusing materials ($n_2>0$) \cite{Christopoulos:2024}. The effect is more pronounced for the first resonance due to the stronger overlap of the mode profile with graphene. In fact, this shift reaches $\sim0.5$~THz (corresponds to a fractional decrease of 12.5\% for the resonant frequency) and  can be used for nonlinearly switching between values of low and high reflection by tuning the input intensity. The contrast (or extinction ratio, ER) can exceed 16~dB when the operating frequency coincides with the first minimum of the black curve in Fig.~\ref{fig:7}(a). The magnitude of the nonlinear shift measured experimentally can be also verified by numerical simulations. The calculations are presented in [Fig.~\ref{fig:7}(b)] and are in good agreement with the experimental measurements.

However, we were not able to identify a definitive feature indicating third harmonic generation. We attribute this to the considerable reflection of the input pulse itself, which retains a non-negligible spectral content even at the high frequencies where we anticipate the third harmonic [cf. Fig.~\ref{fig:DesignEvo}(c)]. Thus, any generated signal cannot be distinguished from the reflected one.  Our assumptions were verified by resorting to time-domain nonlinear simulations. The simulations were conducted using the FDTD commercial software Ansys Lumerical.
%in which graphene is described via the Kubo formula. 
In these simulations we have used three incident source signals with different bandwidths and the results are included in Appendix~C.
%and the same electric field amplitude (8~kV/cm) as shown in Fig.~\ref{fig:time domain}(a-c). 
Briefly, when using a broadband input imitating the actual THz source, a distinct peak of THG cannot be observed in the reflection power spectra. On the other extreme, when we use a narrow source with a bandwidth of 300~GHz, we can identify a clear strong peak of THG. Finally, to verify beyond doubt that the observed peak in the time-domain simulations corresponds to THG we have performed a sweep of the input power and extracted the generated power at the third harmonic frequency. As shown in Appendix~C, the system follows the power scaling law of the third-order process. We therefore concluded that in order to verify the capabilities of the proposed metasurface for efficient THG, the bandwidth of the input pulse should be reduced. However, attempts to do so in the experiment led to significantly limiting the source intensity and, consequently, the efficiency of the nonlinear process. 

%We further investigate the experimental results with frequency domain simulations under continuous wave conditions in commercial software COMSOL Multiphysics using a plane wave x-polarized incident wave with varying electric field amplitude. Graphene is modeled as surface current and its current density is the sum of the  linear and the third-order nonlinear current density corresponding to the Kerr effect \cite{Boyd2020-cq}. More information about the simulations regarding the self-phase modulation is available in Supplementary. The theoretical calculations are presented in [Fig.~\ref{fig:7}(b)] and are in good agreement with the experimental measurements.
 
Thus far, theoretical calculations have been performed under normal incidence. However, in the experimental setup we are forced to deviate from the normal by a small incidence angle (16$^\circ$). Thus, we next studied the effect of the incidence angle on the response of the metasurface, for both TM and TE polarizations. In Appendix~D we include both linear reflection/absorption spectra and the calculated third harmonic generation efficiency. It can be seen  that oblique incidence affects the resonance frequencies and the corresponding absorption peak (reflection dip). As a result, it can affect the THG efficiency, when it deviates considerably from normal incidence for which the metasurface dimensions have been designed. The effect is more detrimental for the TM polarization.

%In Fig.~\ref{fig:Figure S2}, we present the oblique incidence calculations for TM polarization. For this polarization, the magnetic field vector has only a nonzero $H_y$ component; the wavevector has both $k_x$ and $k_z$ components. The angle $\theta$ is the angle between $k$ and $k_z$. The calculated absorption and reflection [Fig.~\ref{fig:Figure S2}(b),(d)] are plotted for different angle values $\theta$, for the optimum metasurface with $r_x = 0.90$ and $r_y = 0.65$.  Finally, we plot the calculated third harmonic generation efficiency and power at third harmonic for oblique incidence as a function of angle $\theta$ [Fig.~\ref{fig:Figure S2}(c)]. The fundamental frequency was chosen to be the one-third of the frequency of the second absorption peak of the resonance.

%\section{Discussion\label{sec:discussion}}

\section{Conclusion\label{sec:conclusion}}
In conclusion, we have proposed a hybrid gold-patch/uniform-graphene nonlinear metasurface  that offers a fabrication-friendly alternative to patterned graphene architectures, while at the same time delivering  highly-nonlinear response. Our theoretical analysis predicts third-harmonic generation efficiencies reaching -15~dB (3.2\%) under continuous-wave excitation at modest input intensities of 0.1~MW/cm$^2$. The metasurface was successfully fabricated via electron-beam lithography and characterized using terahertz time-domain spectroscopy under pulsed excitation. Notably, we experimentally observed pronounced nonlinear spectral shifts of up to 0.5 THz attributed to self-phase modulation, in excellent agreement with simulation results. Altogether, these findings underscore the promise of resonance-engineered graphene-based metasurfaces as a versatile platform for advanced nonlinear photonic functionalities, bridging theoretical feasibility with experimental viability.

\subsection*{Acknowledgment}
We acknowledge the EU Horizon 2020 MSCA project 01007896 (CHARTIST), the Research Council of Finland (Flagship Programme PREIN, decisions no. 368653 and 368652 and research project decisions no. 343393 and 358812), and the Hellenic Foundation for Research and Innovation (H.F.R.I.) under the ``2nd Call for H.F.R.I. Research Projects to support Post-doctoral Researchers'' (Project No. 916, PHOTOSURF) and under the ``2nd Call for H.F.R.I. Research Projects to support Faculty members and Researchers'' (Project Number: 4542).

\appendix
\section{Modeling details}

\subsection{Linear conductivity of graphene}

Graphene is a highly dispersive two-dimensional material; its linear conductivity at the THz regime can be described by a Drude dispersion model:

\begin{equation}
\sigma^{(1)}(\omega) =- \frac{iD}{\pi(\omega-\frac{i}{\tau})},
\end{equation}

\noindent where $D = {q^2\mu_c/\hbar^2}$, $\tau=0.5$~ps is the electron relaxation time, $q$ is the elementary charge, $\mu_\mathrm{c}=0.3$~eV is the chemical potential and $\hbar$ is the reduced Planck’s constant. 

\subsection{Simulating third harmonic generation in CW conditions}

\begin{figure}[h!]
\centering
\includegraphics[width=6cm]{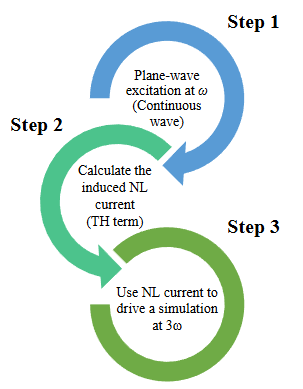}
\caption{Frequency domain simulation procedure for the modeling of third harmonic generation under CW conditions. The simulation is composed of two independent linear simulations at $\omega_\mathrm{FF}$ and at $3\omega_\mathrm{FF}$.}
\label{fig:simulation steps}
\end{figure}

Third-harmonic generation (THG) is a third-order nonlinear phenomenon in which three photons at the fundamental frequency, $\omega_\mathrm{FF}$,  produce a single photon at the third harmonic frequency, $3\omega_\mathrm{FF}$. 
Nonlinear processes are best modeled with time-domain simulations. However, frequency-domain simulations can be also used when discussing continuous wave (CW) illumination. For the case of third harmonic generation (THG), we can use two linear simulations \cite{Christopoulos2018, Jin2017}, as shown in Fig. \ref{fig:simulation steps}. The first simulation concerns plane-wave scattering at the fundamental frequency, $\omega_{FF}$. The linear conductivity of graphene is used and a linear current is induced on graphene. 
The second step is to calculate the induced (linear and nonlinear) current density:
\begin{equation}
{\textbf{J}_{3\omega}= \sigma^{(1)}\mathbf{E}_{||} e^{j3\omega t}+
\sigma^{(3)}\mathbf{E}_{||}\cdot
\mathbf{E}_{||}\cdot \mathbf{E}_{||} e^{j3\omega t}}
\end{equation} 
using graphene’s third-order conductivity formula by Cheng et al. \cite{Cheng2014}, simplified for the case of low (THz and microwave) frequencies ($\hbar\omega_\mathrm{{FF}}/\mu_\mathrm{c}\ll 1$) \cite{Mikhailov2007, Mikhailov2008}:
\begin{equation}
{\sigma^{(3)}(\omega)=\frac{i\sigma_0(\hbar u_\mathrm{F}q)^2}{2 \pi \mu_\mathrm{c} ( \hbar \omega_\mathrm{{FF}})^3}}.
\label{eq:sigma3}
\end {equation}
In Eq.~\eqref{eq:sigma3}, $\sigma_0 = q^2/4\hbar$ and $u_\mathrm{F}=10^6$~m/s is the Fermi velocity. 
Having calculated the nonlinear current density, the final step is to perform a second simulation (without plane wave excitation) letting the induced nonlinear current radiate at the third harmonic frequency $\omega_\mathrm{TH}=3\omega_\mathrm{FF}$. The radiated field is collected and the THG conversion efficiency (CE) is calculated as the ratio $P_\mathrm{TH}/P_\mathrm{FF}$ or $I_\mathrm{TH}/I_\mathrm{FF}$, where $P_\mathrm{TH}$ ($I_\mathrm{TH}$) is the outward radiated power (intensity) and $P_\mathrm{FF}$ ($I_\mathrm{FF}$) is the injected power (intensity) at the fundamental frequency. 

%THG in the nonlinear metasurface under study is simulated using full-wave electromagnetic simulations with COMSOL Multiphysics commercial software (Finite Element Method).  To calculate the nonlinear response of the metasurface under continuous wave (CW) illumination, we decouple the nonlinear problem into two linear frequency-domain simulations, at the fundamental and third-harmonic frequency \cite{Jin2017, Christopoulos2018}. This is generally not possible for pulsed excitation. 

%The first simulation is conducted at the fundamental frequency, using the graphene's linear surface conductivity, with plane wave excitation. Graphene is naturally modeled as an infinitesimally thin material via a surface current boundary condition. 

\subsection{Simulating self-phase modulation in CW conditions}

Self-phase modulation (SPM) is also a third-order nonlinear phenomenon in which the material property at the operating  frequency (surface conductivity or refractive index) changes due to the optical Kerr effect. For CW conditions, SPM can be simulated in the frequency domain. Unlike THG where two frequencies are involved, here a single simulation step suffices. The induced linear and nonlinear current density in this case is:

\begin{equation}
{\textbf{J}_{\omega}= \sigma^{(1)}\mathbf{E}_{||} e^{j\omega t}+
\sigma^{(3)}\mathbf{E}_{||}\cdot
\mathbf{E}_{||}^{*}\cdot \mathbf{E}_{||} e^{j\omega t}}
\end{equation}

\section{Parametric study for investigating the double resonant condition}

\begin{figure}[h!]
\centering
\includegraphics[width=15cm]{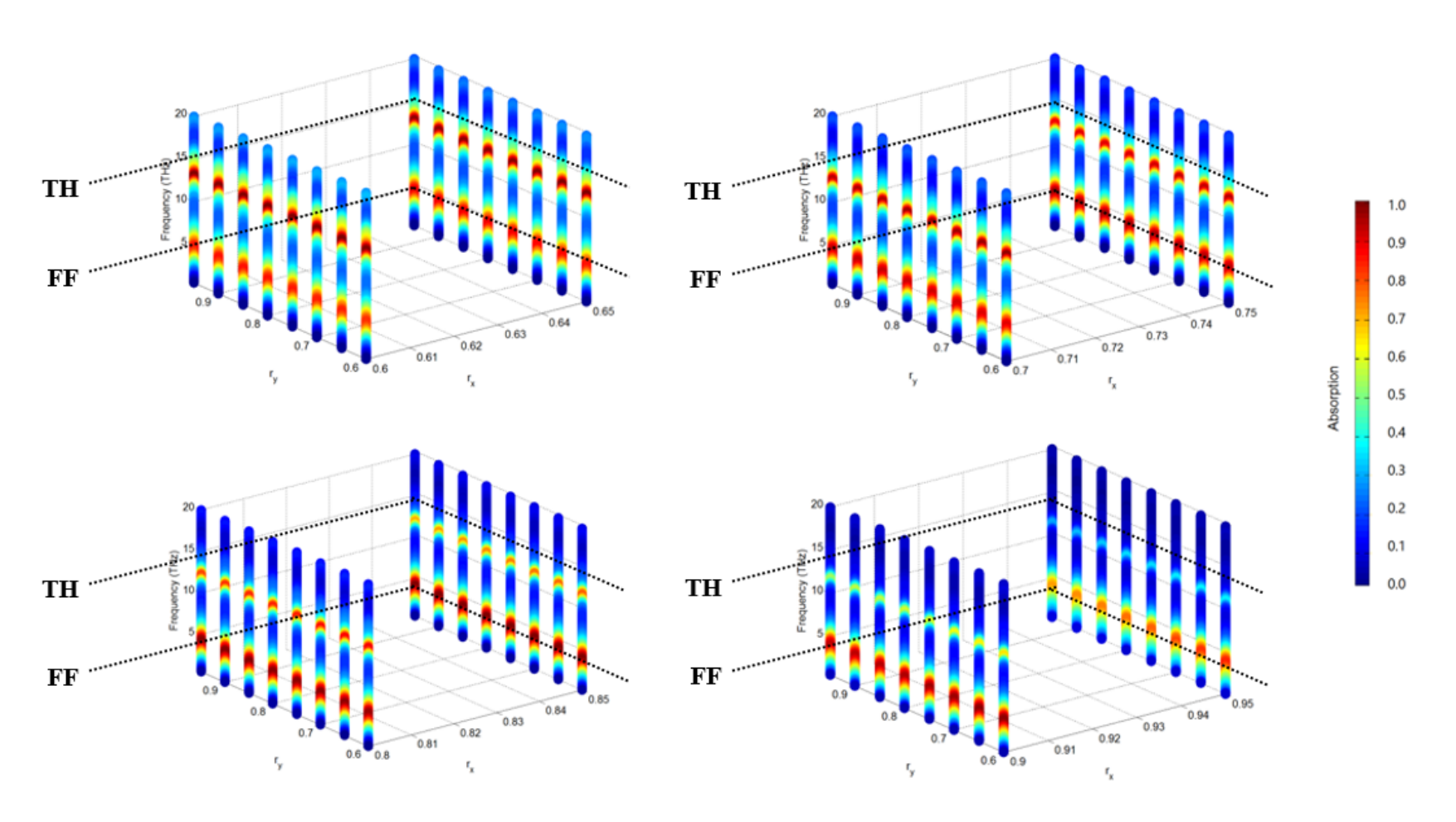}
\caption{Calculated absorption of the hybrid graphene metasurface, as a function of the frequency and meta-atom size: $ r_x =  a_x/a $ and $r_y = a_y/a$. The $a$ parameter denotes the lattice constant and  equals $5.6$~$\mu$m. The absorption peaks of the two resonances are  aligned at the $\omega$ - 3$\omega$ configuration for $r_x = 0.9$ and $r_y = 0.65$.\label{fig:Absorption}}
\end{figure}

In Fig.~\ref{fig:Absorption} we present the calculated absorption of the hybrid gold-graphene metasurface as a function of the frequency and the size of meta-atom, which varies through the parameters $ r_x =  a/a_x $ and $r_y = a / a_y$. The $a$ parameter denotes the lattice constant and  equals $5.6$~$\mu$m. The two resonances of the metasurface are  aligned at the $\omega$ - 3$\omega$ configuration for $r_x = 0.9$ and $r_y = 0.65$.

%\section{Step-by-step step description of the fabrication process}

%The detailed description of the experimental procedure is presented in Fig.~\ref{fig: Exp_procedure}.
%\begin{figure}[h!]
%\centering
%\includegraphics[width=9cm]{Figures/Exp_procedure.pdf}
%\caption{Step-by-step step description of the fabrication process.\label{fig: Exp_procedure}}
%\end{figure}

\clearpage

\section{Time-domain nonlinear simulations with pulsed input}

\begin{figure}[htb]
\centering
\includegraphics[width=12cm]{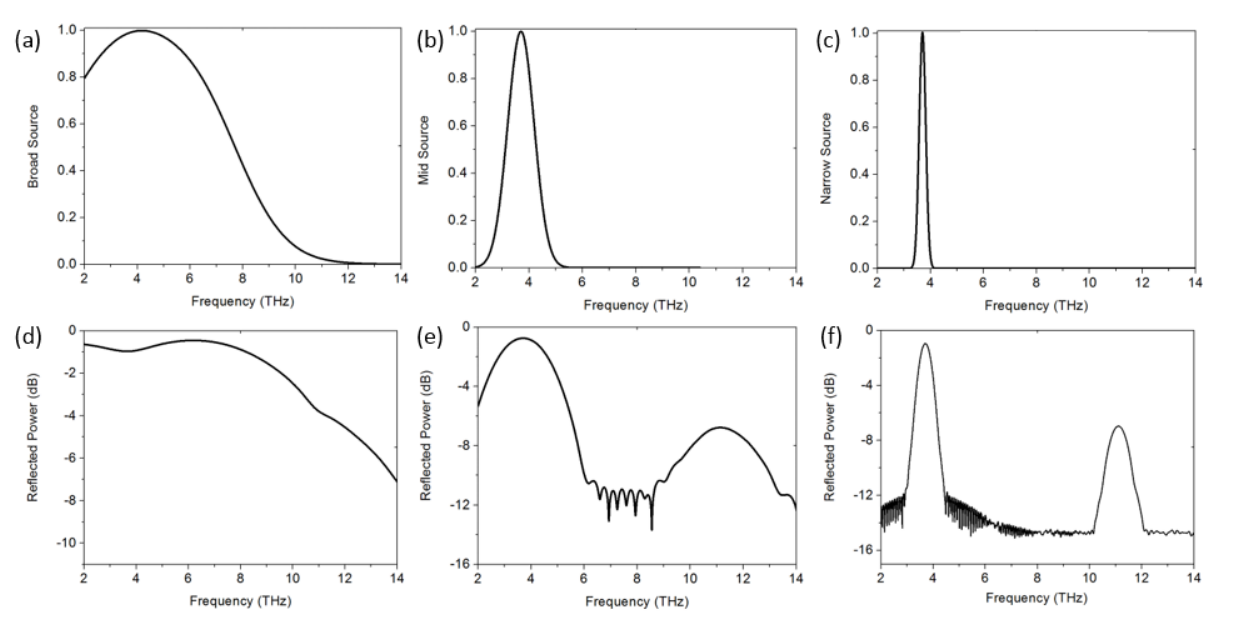}
\caption{Time-domain simulations for an incident electric field amplitude 8~kV/cm and different pulse bandwidths. (a-c)~Spectra of the three different pulsed sources with progressively narrower bandwidth. (d-f)~Corresponding reflection power spectra.\label{fig: Time_domain}}
\end{figure}

The time-domain nonlinear simulations (ANSYS Lumerical) are presented in Fig.~\ref{fig: Time_domain}. In these simulations, we used three incident source signals with different bandwidths and the same electric field amplitude (8 kV/cm) as shown in Fig.~\ref{fig: Time_domain}(a-c). In more detail, when using a broadband input imitating the actual THz source, such as the one
shown in Fig.~\ref{fig: Time_domain}(a), a distinct peak of THG cannot be observed in the reflection power spectra [Fig.~\ref{fig: Time_domain}(d)]. On the other extreme, when we use a narrow source with a bandwidth of 300~GHz [Fig.~\ref{fig: Time_domain}(c)], we can identify a clear strong peak of THG [Fig.~\ref{fig: Time_domain}(f)].

\clearpage

\begin{figure}[htb]
\centering
\includegraphics[width=7cm]{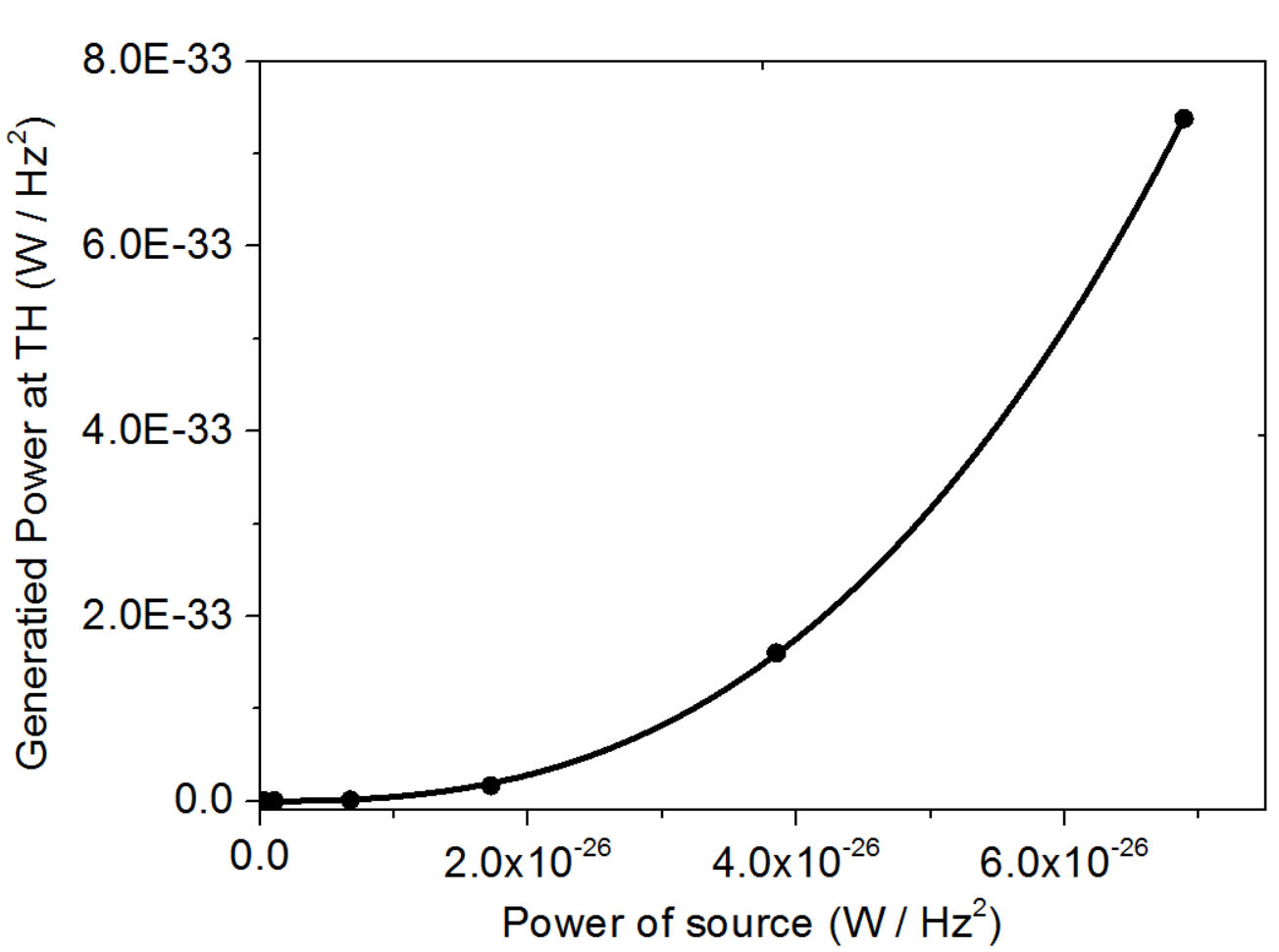}
\caption{Generated third harmonic intensity as a function of the input source intensity. The curve follows the power scaling law of the third-order process.  \label{fig: Power}}
\end{figure}

Finally, to verify beyond doubt that the observed peak near 12~THz in Fig.~\ref{fig: Time_domain} corresponds to THG we have
performed a sweep of the input power and extracted the generated power at the third harmonic
frequency. As shown in Fig.~\ref{fig: Power}, the system follows the power scaling law of the third-order process.

\clearpage

\section{Oblique incidence calculations}

\begin{figure}[h!]
\centering
\includegraphics[width=10cm]{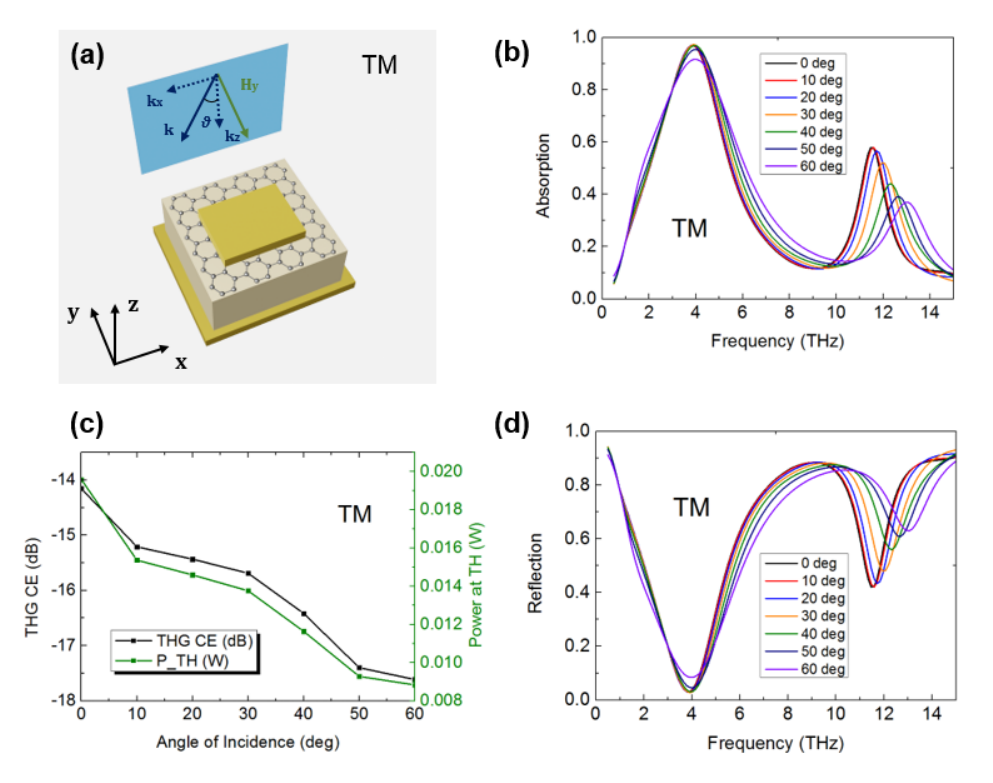}
\caption{Oblique incidence calculations for the hybrid gold-graphene metasurface for TM polarization inside the $xz$ plane. (a) For this polarization, the magnetic field vector has only the $H_y$ component; the wavevector has both $k_x$ and $k_z$ components. The angle $\theta$ is the angle between $k$ and $k_z$ and calculations refer to the optimum metasurface with $r_x = 0.90$ and $r_y = 0.65$ (b),(d) Calculated absorption and reflection, for different values of the angle $\theta$. (c)~Calculated third harmonic generation efficiency and power at third harmonic for oblique incidence as a function of angle $\theta$. The fundamental frequency was chosen to be one-third of the absorption peak of the second resonance.  For higher  angles, the 1:3 resonant frequency configuration is disturbed and THG conversion efficiency deteriorates.\label{fig: Oblique_TM}}
\end{figure}

In Fig.~\ref{fig: Oblique_TM} we present the oblique incidence calculations for TM polarization inside the $xz$ plane. For this polarization, the magnetic field has only the $H_y$ component while the wavevector has both $k_x$ and $k_z$ components. The angle $\theta$ is the angle between $k$ and $k_z$. The calculated absorption and reflection are plotted for different angle values $\theta$, for the optimum metasurface with $r_x = 0.90$ and $r_y = 0.65$. Finally, we plot the calculated third harmonic generation efficiency and power at third harmonic for oblique incidence as a function of angle $\theta$. The fundamental frequency was chosen to be one-third of the absorption peak of the second resonance.

\clearpage

\begin{figure}[h!]
\centering
\includegraphics[width=10cm]{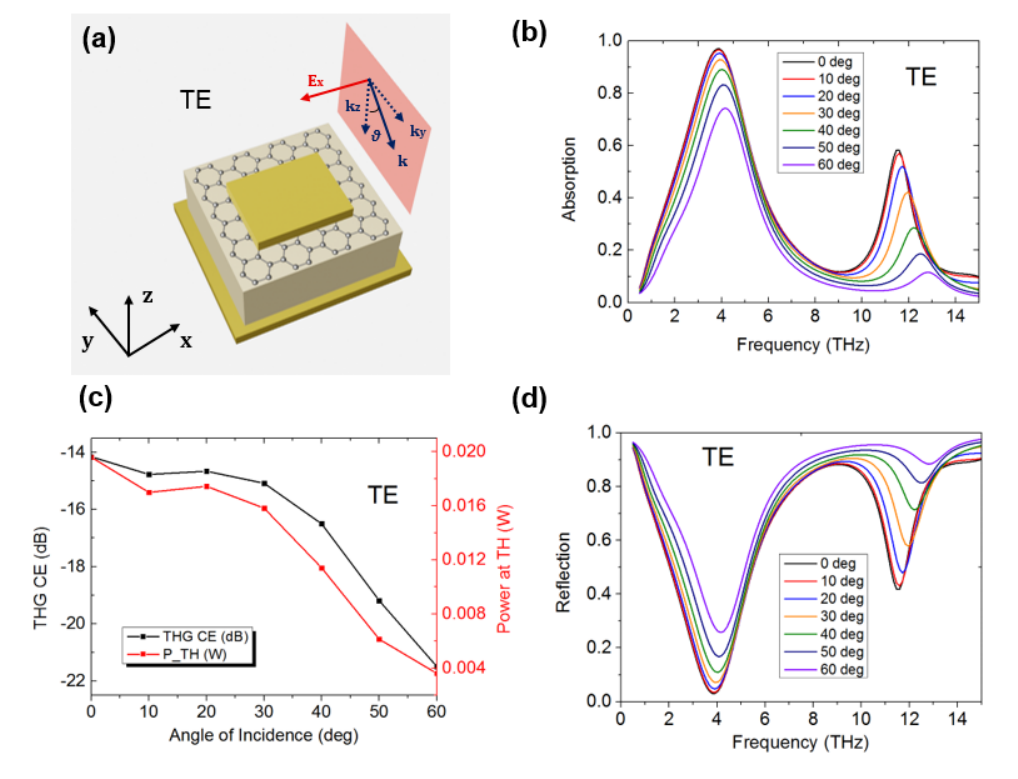}
\caption{Oblique incidence calculations for the hybrid gold-graphene metasurface for TE polarization inside the $yz$ plane. (a)~The electric field has a single nonzero component ($E_x$), whereas the wavevector has both $k_y$ and $k_z$ components. (b),(d)~Calculated absorption and reflection, for different values of the angle $\theta$, for the optimum metasurface with $r_x = 0.90$ and $r_y = 0.65$. (c)~Calculated third harmonic generation efficiency and power at third harmonic for oblique incidence as a function of angle $\theta$. The fundamental frequency was chosen to be the one third of the absorption peak of the second resonance. For higher  angles, the 1:3 resonant frequency configuration is disturbed and THG conversion efficiency deteriorates.\label{fig:Oblique_TE}}
\end{figure}

In Fig.~\ref{fig:Oblique_TE} we present oblique incidence calculations for TE polarization inside the $yz$ plane. In this case, the electric field has a single nonzero component ($E_x$). The wavevector has both $k_y$ and $k_z$ components and the angle $\theta$ is measured between $k$ and $k_z$. Again, we see a deterioration of the conversion efficiency as the incident angle increases. This is because the frequency and linewidth of the second resonance are disturbed as the  incidence angle increases. If we wanted to operate more efficiently for a specific oblique incidence angle, we should repeat the design process for this particular angle. 

\bibliographystyle{unsrt}
%\bibliography{literature}

\end{document}